\documentclass[prl,twocolumn,showpacs,amsmath,amssymb]{revtex4}

\usepackage{amsmath}
\usepackage{graphicx}
\usepackage{epsfig}

\newcommand{\be}{\begin{equation}}
\newcommand{\ee}{\end{equation}}
\newcommand{\bma}{\begin{displaymath}}
\newcommand{\ema}{\end{displaymath}}

\begin{document}

\title{Rotating electrons in quantum dots: Quantum Hall liquid in the classical limit}

\author{J.-P. Nikkarila and M. Manninen}

\affiliation{\sl NanoScience Center, Department of Physics,
FIN-40014 University of Jyv\"askyl\"a, Finland}

 
\date{today}

\begin{abstract} 

We solve the problem of a few electrons in a
two-dimensional harmonic confinement using quantum mechanical 
exact diagonalization technique, on one hand, and classical
mechanics, on the other hand. 
The quantitative agreement between the results of these two 
calculations suggests that, at low filling factors, all the 
low energy excitations of quantum Hall liquid are classical
vibrations of localized electrons. 
The Coriolis force plays a dominant role 
in determining the classical vibration frequencies.

\end{abstract}

\maketitle


Spectroscopic studies of semiconductor quantum dots 
have shown that in zero magnetic field they behave like 'artificial
atoms' showing electronic
shell structure and obeying Hund's rule\cite{tarucha1996}, while in a strong
magnetic field they have properties of quantum Hall liquids\cite{mceuen1991,oosterkamp1999}.
The integer quantum Hall effect corresponds in a quantum dot the
formation of so-called maximum density droplet\cite{macdonald1993} of polarized
electrons.
For higher magnetic fields the research has mainly been theoretical,
showing formation of quantum
liquids related to the 
fractional quantum Hall effect (QHE)\cite{laughlin1983,maksym1990,wojs1997}.
Although the quantum dot is finite and has a surface, its
electronic structure at high magnetic fields shows
properties of the QHE, which is usually studied using
geometries with periodic boundary conditions, i.e. the surface
of a sphere\cite{he1994} or a torus\cite{pfannkuche1997}.

The generic model of a semiconductor quantum dot is very simple:
Electrons interacting with Coulomb interaction are confined in a
two-dimensional harmonic potential. Due to the circular symmetry,
the angular momentum is a good quantum number, and due to the harmonic
confinement the center-of-mass motion exactly separates out\cite{chakraborty1999,reimann2002} 
from the internal motion of the particles.
The wave functions of non-interacting electrons are simple 
(with or without an external magnetic field).
Moreover, in a strong magnetic field, once the electrons are
polarized, the only effect of the magnetic field is to
put the electron system in rotation, i.e. increase the
total angular momentum. For small number of electrons the 
many-particle problem can be solved numerically exactly.
It is then not surprising that the quantum dot geometry 
has been extensively used for testing different models of the
fractional quantum Hall liquid  (e.g. the Laughlin
wave function\cite{laughlin1983} and Jain construction\cite{jain1989}).

The purpose of this letter is to show that the quantum mechanical
many-particle spectrum of strongly correlated electrons
of the QHE system can be {\it quantitatively} determined
by solving the Newton equation of motion. To this end 
we solve, for small number of electrons,
the excitation spectra using many-particle 
quantum mechanics and compare the results to those
determined by quantizing the rotational and vibrational
modes calculated in a rotational frame for a Wigner molecule
using classical mechnanics.

We assume a generic model of a polarized (or spin-less) 
electrons interacting with 
unscreened Coulomb interaction in a two-dimensional harmonic
potential. The Hamiltonian is
\be
H=-\frac{\hbar^2}{2m}\sum_{i}^N \nabla_i^2 
+\sum_i^N \frac{1}{2}m\omega_0^2 r_i^2
+\sum_{i<j}^N \frac{e^2}{4\pi\epsilon_0\vert {\bf r}_i-{\bf r}_j\vert}
\label{hamiltonian}
\ee
where $N$ is the number of particles, $m$ the electron mass, 
${\bf r}=(x,y)$ a two-dimensional position vector, and $\omega_0$ the 
oscillation frequency of the confining potential.
We solve the many-particle problem for a fixed angular momentum
using the single particle basis of the lowest Landau level (LLL):
\be
\psi_\ell(r,\phi)=A_\ell r^\ell e^{-r^2/4}e^{i\ell\phi},
\label{spstate}
\ee
where $\ell$ is the single-particle angular momentum
and $A_\ell$ a normalization factor. 
We are interested in solutions for large total angular momenta.
In this case the restriction of the basis in the LLL is a good 
approximation\cite{manninen2001}. 
Note that we do not explicitly include magnetic field 
in our calculations: 
The only effects the magnetic field would have, 
were to polarize the electron
gas due to the Zeeman effect and to increase the total angular
momentum. We present our results in terms of angular momentum.
The total angular momentum can be related to the filling
factor of electrons with the relation
$\nu=N(N-1)/2L$, where $N$ is the number of electrons and $L$ the
total angular momentum.

We solve the many-particle Schr\"odinger equation using
the straightforward CI technique. 
The many-particle states are linear combinations of Slater
determinants of single particle configurations. The number of
configurations needed increases fast with $N$ and $L$, but
for the results shown here a complete basis in the LLL
could still be used. For determining the Coulomb matrix elements 
we used the method of Ref. \cite{stone1991}

Classical electrons in a 2D harmonic trap form Wigner molecules,
where the electrons arrange in consecutive circles\cite{bolton1993,bedanov1994}.
The elementary excitations of the classical system consist of the
center of mass vibrations with angular frequency $\omega_0$,
internal vibrations which can be solved from the dynamical matrix,
and rigid rotations of the Wigner molecule. Since we are interested in the
excitations of a rotating system (with fixed angular momentum)
we have to solve the internal vibrations in a rotating frame\cite{matulis2005}.
Taking into account the Coriolis force
turns out to be crucial in the case of electrons in a harmonic trap
(although its effect in real molecules is small).

\begin{figure}[h]
\includegraphics[width=\columnwidth]{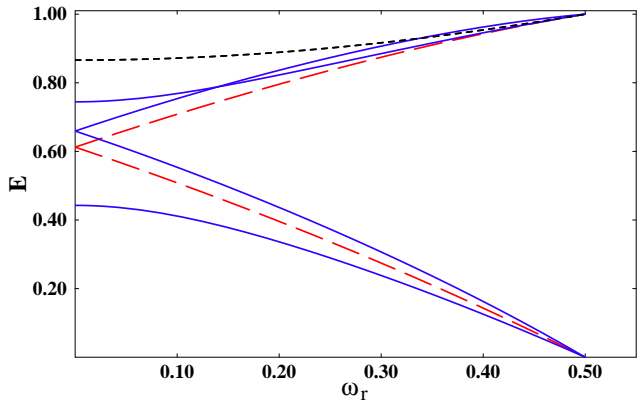}
\caption{Classical vibrational frequencies as a function
of the angular velocity of the rotating Wigner molecule
with 2, 3, and four electrons. The breathing mode 
(only mode for two electrons) is shown as a dotted line.
The solid (dashed) lines show other modes for four (three) electrons
(in atomic units with $\omega_0=0.5$).
}
\end{figure}

The equilibrium positions of the electrons depend on the 
angular velocity $\omega_r$ or angular momentum $L=I\omega_r$
of the molecule ($I$ is the moment of inertia $I=\sum mr_i^2$),
and they can be solved by minimizing the classical energy
\be 
E_{\rm cl}^0(L)=\frac{1}{2}m\omega_0\sum_i^N r_i^2
+\sum_{i<j}\frac{e^2}{4\pi\epsilon_0\vert{\bf r}_i-{\bf r}_j\vert}
+\frac{L^2}{2m\sum r_i^2}.
\ee
The eigenfrequencies of the vibrations can then be 
solved analytically from the equations of motion of the rotating frame
(by linearizing the equations around the equilibrium positions
of electrons).
Results for classical vibrational frequencies for 
2, 3 and 4 electrons are shown in Fig. 1.
The eigenfrequencies are shown 
as a function of the angular velocity of the rotation $\omega_r$.
We notice that all frequencies approach to either 0 or $2\omega_0$.
In all calculations shown in this letter we have used
atomic units ($m=\hbar=e=a_0=1$) and chosen $\omega_0=0.5$.
We should note that in the LLL the results are independent
of $\omega_0$, apart from the energy scale.

For each particle number the highest mode corresponds to 
the breathing mode where the molecule expands and shrinks 
without changing its shape. In a rotating frame, however,
each electron moves along an ellipse around its 
equilibrium position. In the case of two electrons this mode is 
naturally the only vibrational mode. For 2, 3 and 4
particles the breathing mode energy is the same
and its dependence of the angular velocity is similar
$\omega_{\rm bm}=\sqrt{3\omega_0+\omega_r}$
(this is general result for a Wigner molecule where
the electrons form one single ring).

In the case of 3 and 4 electrons the non-rotating system
has degenerate vibrational modes.
However, when the system is put in rotation the
degeneracy will split. 
In the case of three electrons the ground state is an
equilateral triangle. In the doubly degenerate
vibrational mode the triangle stays isosceles, the base 
stretching and shrinking. In the non-rotating system 
we can choose any of the sides to be the base, but only
two of them linearly independently. However, we can also
choose a linear combination of the two modes in such a way 
that the base circulates around. In this {\it pseudo-rotation}
the electrons move in circles around their equilibrium positions
(as indicated in Fig. 4). The degeneracy then follows from the two
possible direction of the pseudo-rotation. If the molecule is put
in rotation, the Coriolis force splits
this degeneracy. Similar analysis can be made for the
degenerate mode in the case of four electrons.
It is interesting to note that in the case of sodium trimer,
where the Jahn-teller effect opens the equilateral 
triangle, a related pseudo-rotation have been observed\cite{delacretaz1986}.

\begin{figure}[h]
\includegraphics[width=0.9\columnwidth]{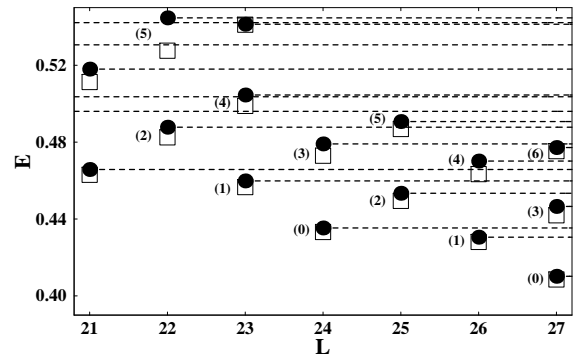}
\caption{Many-particle energy spectrum for three electrons.
The interaction energy is shown as a function of the angular momentum.
The black dots are results of the Schr\"odinger equation and the
open squares the results from the classical model. The center
of mass excitations are not shown as points but are indicated as dashed lines.
The numbers indicate the order of the vibrational state ($n_1$).
}
\end{figure}
\begin{figure}[h]
\includegraphics[width=0.9\columnwidth]{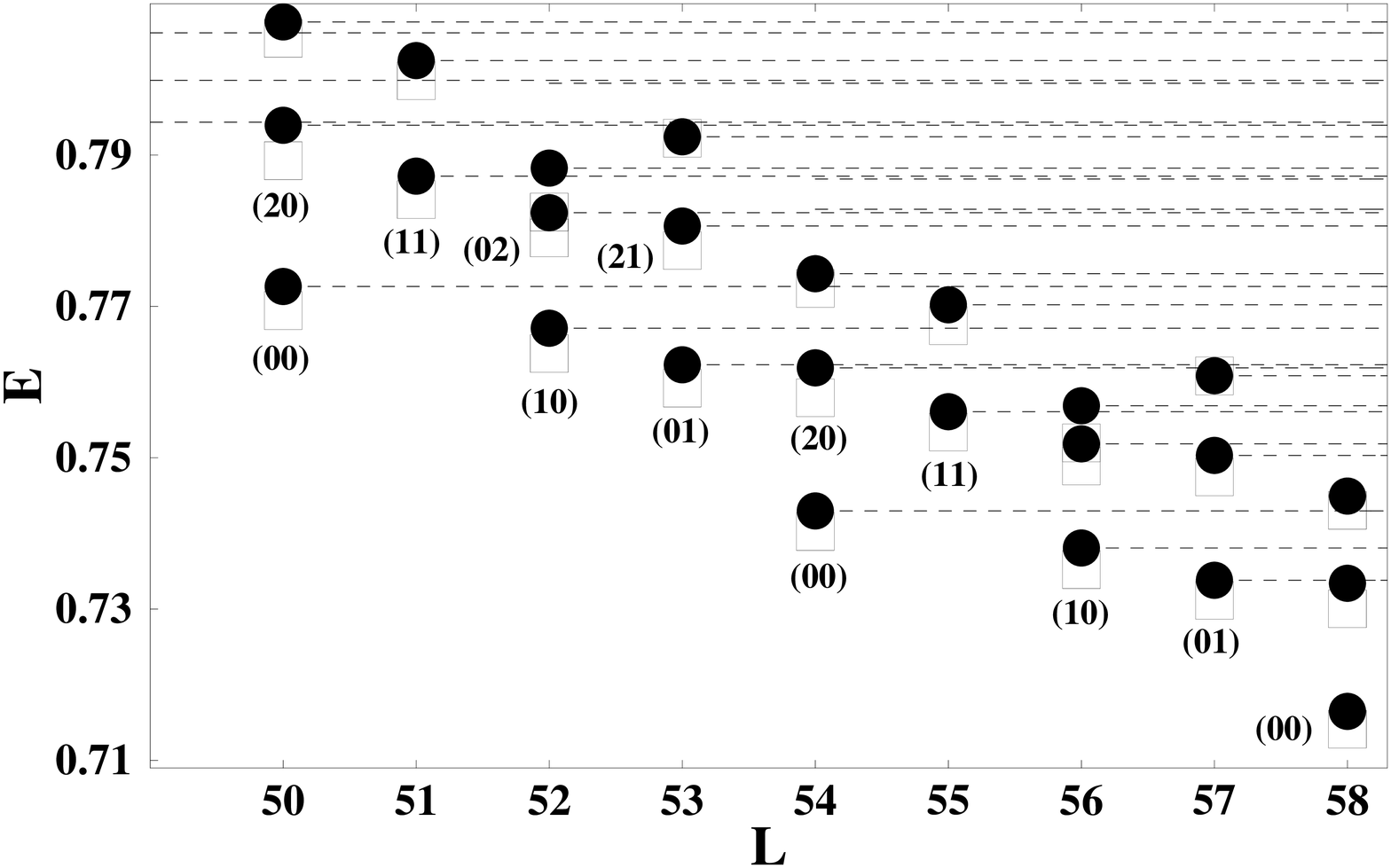}
\caption{Many-particle energy spectrum for four electrons.
The interaction energy is shows as a function of the angular momentum.
The black dots are results of the Schr\"odinger equation and the
open squares the results from the classical model. The center
of mass excitations are not shown as points but are indicated as dashed lines.
The numbers indicate the vibrational state  ($n_1,n_2$).
}
\end{figure}

Having the classical vibrational frequencies in hand, we follow the 
normal assumptions in quantization of phonons\cite{ashcroft1976} and
get an estimate for the total energy as
\be
E_{\rm cl}=E_{\rm cl}^0+\sum_k \hbar\omega_k(n_k+\frac{1}{2})+
\hbar\omega_0(n_{\rm 0}+1),
\ee
where $\omega_k$ are all the vibrational frequencies determined in the 
rotating frame and $n_k=0,~1,~2,\cdots$, and the last term
corresponds to the center of mass excitations.
These energies can now be compared with those obtained by solving the 
many-particle Schr\"odinger equation. 
However, we still have to note that the quantum state for fermions
has to be antisymmetric. This means that not all combinations of $L$ and $n_k$
are allowed (but any center-of-mass excitation is always possible).
The symmetry analysis can be made with help of group 
theory\cite{tinkham1964,koskinen2001,koskinen2002} and it shows, for example,
that without any vibrational modes only every second angular momentum
is allowed for two particles, every third for the triangle of three 
particles and every fourth for the square of four particles. 
The lowest energy of any other angular momentum
value must have either a center of mass excitation or an internal
vibration in addition of the internal rotation. This fact leads to
the well known oscillations of the lowest energy as a function of the 
angular momentum\cite{wojs1997,reimann2002,reimann2006}.

Figures 2 and 3 show the lowest energies for three and four particles
obtained using two different ways: Solving exactly the many-particle 
Schr\"odinger equation and by solving the vibrational modes using
classical mechanics and determining the energy from Eq. (2).
The figures show the 'interaction energy', the energy difference between
the interacting and non-interacting electrons,
$E-(N+L)\hbar\omega_)$, where $E$ is the solution of the 
Schr':odinger equation of of Eq. (4).
In this presentation
the center-of-mass excitations become horizontal (the center-of-mass motion
is independent of the interactions) and they are only indicated as 
dashed lines. The results show that spectra determined by quantizing
the classical vibrations agree very well with those from the
full quantum mechanical calculations. 
The error in the total interaction energy is less than 1 \%.
We note in passing that in the
case of the frequently studied system of 
two electrons\cite{yannouleas2000,helle2005} 
the low energy spectrum only consists of 
rigid rotations and center-of-mass excitations, since the only 
vibrations mode is high in energy.

\begin{figure}[h]
\includegraphics[width=0.9\columnwidth]{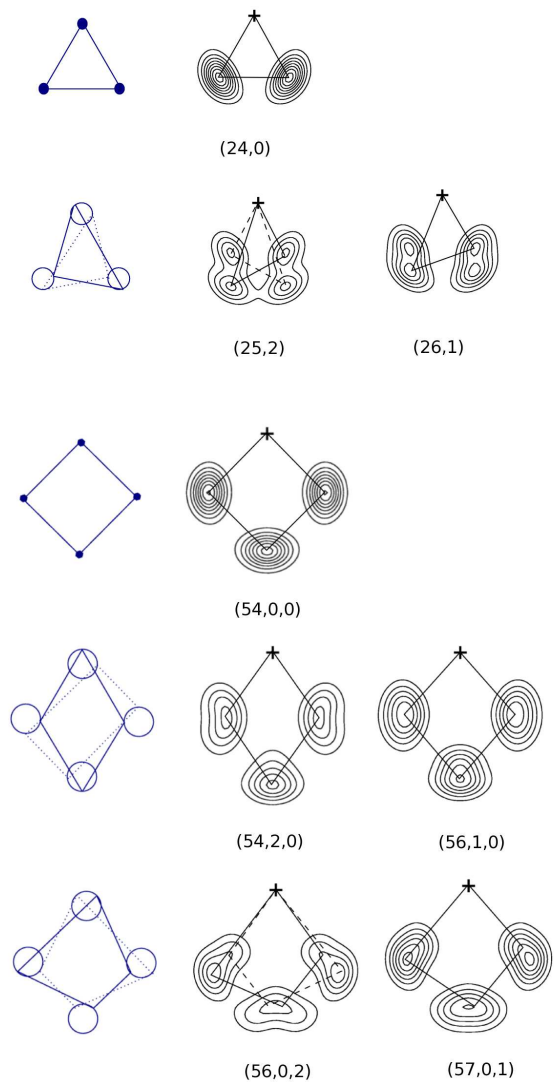}
\caption{The left column shows the classical geometries and 
schematically the low energy vibrational 
modes (pseudo-rotations) for three
and four electrons. The two other columns show contour 
plots of the pair correlation
functions from the quantum mechanical calculation. 
The reference point is shown as a cross. The relation
of the pair correlation functions to the classical extreme
geometries is indicated with solid and dashed lines.
The numbers in parentheses are ($L,n_1,n_2$) indicate the
angular momentum and the vibrational state.
}
\end{figure}

In the case of three electrons the purely rotational state
can occur at angular momenta $L=3n$, where $n$ is an integer.
For $L=3n+1$ the lowest energy state has a center of mass excitation
(the point not shown) and for $L=3n+2$ the lowest energy state has
vibrational mode $\omega_1$. The excited states for any angular
momentum show simple systematics consisting of center of mass 
excitations and multiplets of $\omega_1$, as indicated in Fig. 3.

In the case of four particles the energy spectrum is more complicated 
due to the two low energy vibrational modes. Nevertheless, Fig. 3
shows that also in this case the spectrum consists of periodic 
sequences, but the length of the period is now four. 

In Figures 2 and 3 we have shown the spectra for large angular momenta,
where the quantization of the classical energies give quantitatively
accurately the energy differences. The angular momentum $L=54$
for four particles and $L=27$ for three particles
corresponds to to the filling factor $\nu=1/9$ of a quantum Hall
liquid, where the electrons are expected to be localized to 
a Wigner crystal. However, the excitation spectrum, when plotted as
in Figs. 2 and 3 are {\it qualitatively} similar already from
angular momenta corresponding to filling factor $\nu=1/3$.

Finally, we want to show that the pair correlation functions 
determined from the full quantum mechanical calculation
are consistent with the vibrational modes determined from
classical mechanics. Figure 4 shows schematically
the classical vibrational modes in the the rotational frame
compared to the pair correlation functions determined for
different states shown in the spectra in Figs. 2 and 3.
The pair correlation functions for the purely rotational states
shows clearly the localization of the electrons, and their shape is
in agreement with the semiclassical model ov Matulis and Anisimovas\cite{matulis2005}.
However, the interpretation of the vibrational states is not so simple, because 
the structure of the pair correlation function depends on the reference point. 
In Fig. 4 the reference electron is chosen to be at 
the extreme point of the classical motion. In this case we can
identify the classical extreme positions as shown as triangles and
quandrangles drawn on top of the pair correlation functions. Note that
the pair correlation function can capture the classical geometry in two 
different positions.


In conclusion, we have determined the classical rotational and vibrational modes
of electrons in a two-dimensional harmonic trap using a rotational
frame. Results of the quantized energies are compared with numerically
exact solutions of the many-particle Schr\"odinger equation, and they
indicate clearly that when the angular momentum of the system
increases to the region which corresponds to the filling factor of 1/9
of the quantum Hall liquid, the whole low-energy excitation spectrum
can be quantitatively described with the classical rotations and
vibrations. The pair-correlation functions are consistent with the
classically determined vibrational modes.

Acknowledgments. 
We would like to thank Ben Mottelson, Matti Koskinen and 
Stephanie Reimann for valuable discussions. This work 
was supported by the Academy of Finland.

\end{document}